\documentclass[pra,
twocolumn,
hyperref,
showkeys
]{revtex4-1}

\usepackage{amsmath}
\usepackage{hyperref}
\usepackage{bm}
\usepackage{dsfont}
\usepackage{amsthm}
\usepackage{verbatim} 
\usepackage{qcircuit}
\usepackage{circuitikz}
\usepackage{tabularx}
\usepackage{amssymb}
\usepackage[size=small]{caption}
\usepackage{algcompatible}
\usepackage{newfloat}
\usepackage{etoolbox}
\usepackage{xcolor}
\usepackage{soul}

\AtBeginEnvironment{algorithm}{\noindent\hrulefill\par\nobreak\vskip-5pt}
\usepackage{newfloat}

\DeclareFloatingEnvironment[
    fileext=loa,
    listname=List of Algorithms,
    name=ALGORITHM,
    placement=tbhp,
]{algorithm}
\DeclareCaptionFormat{algorithms}{\vskip-15pt\hrulefill\par#1#2#3\vskip-6pt\hrulefill}
\definecolor{burgundy}{rgb}{0.5, 0.0, 0.13}
\hypersetup{colorlinks,linkcolor={blue},citecolor={blue},urlcolor={burgundy}}  
\newcommand{\bra}[1]{\mbox{$\langle #1|$}}
\newcommand{\ket}[1]{\mbox{$|#1\rangle$}}
\newcommand{\braket}[2]{\mbox{$\langle #1|#2\rangle$}}
\newcommand{\ketbra}[2]{\mbox{$|#1\rangle\langle #2|$}}

\newcommand{\myTriangleBlack}[4] 
{\draw [rotate around={#3: (#1+.35, #2)}, fill=black] (#1, #2) +(0,.4) -- +(0,-.4) -- +(.7,0) -- +(0,.4) (#1+.25, #2) node[white] {#4}}

\begin{document}

\author{Andrey Kardashin}
\email[]{andrey.kardashin@skoltech.ru}
\author{Alexey Uvarov}
\email[]{alexey.uvarov@skoltech.ru}
\author{Jacob Biamonte}
\email[]{j.biamonte@skoltech.ru}
\homepage[]{http://quantum.skoltech.ru}
\affiliation{Deep Quantum Laboratory, Skolkovo Institute of Science and Technology, 3 Nobel Street, Moscow, Russia 121205}

\title{Quantum Machine Learning Tensor Network States}

\date{\today}

\begin{abstract}
 Tensor network algorithms seek to minimize correlations to compress the classical data representing quantum states.
 Tensor network algorithms and similar tools---called tensor network methods---form the backbone of modern numerical methods used to simulate many-body physics and have a further range of applications in machine learning.
 Finding and contracting tensor network states is a computational task which quantum computers might be used to accelerate.  
 We present a quantum algorithm which returns a classical description of a rank-$r$ tensor network state satisfying an area law and approximating an eigenvector given black-box access to a unitary matrix. Our work creates a bridge between several contemporary approaches, including tensor networks, the variational quantum eigensolver (VQE), quantum approximate optimization (QAOA), and quantum computation.  
\end{abstract}

\keywords{quantum algorithms, VQE, QAOA, quantum machine learning, optimisation; tensor networks, matrix product states; linear algebra, eigenvectors}

\maketitle

\section*{Introduction}

Tensor network methods provide the contemporary state-of-the-art in the classical simulation of quantum systems.  A range of numerical and analytical tools have now emerged, including tensor network algorithms to simulate quantum systems classically---these algorithms are based in part on powerful insights related to the area law \cite{2017arXiv170301302B, 2014AnPhy.349..117O, Vidal2010, MPSreview08, TNSreview09, 2011AnPhy.326...96S,  2014EPJB...87..280O,2013arXiv1308.3318E, 2011JSP...145..891E}.  The area law places bounds on quantum entanglement that a many-body system can generate, which translates directly to the amount of memory required to store a given quantum state, see e.g.~\cite{2013arXiv1308.3318E}. 

The leading classical methods to simulate random circuits for quantum computational supremacy demonstrations are also based on tensor network contractions. Additionally, classical machine learning has been merged with matrix product states and other tensor network methods \cite{MAL-059, MAL-067, 1751-8121-51-13-135301, Huggins_2019,2017arXiv171004833L}.  How might quantum computers accelerate tensor network algorithms?  

Although tensor network tools have traditionally been developed to simulate quantum systems classically, we propose a quantum algorithm to approximate an eigenvector of a unitary matrix with bounded rank tensor network states.
The algorithm works given only black-box access to a unitary matrix. In general, tensor network contraction can simulate any quantum computation.  

We focus on 1D chains of tensors (matrix product states) due to some associated analytical simplifications: indeed, matrix product states can be approximated classically which offers an attractive gold standard to compare the quantum algorithm against.  The general framework we develop applies equally well to 2D and e.g.~sparse networks (projected entangled pair states, etc.). However, an early merger between these topics is better situated to focus on 1D.

Even in 1D, tensor networks offer certain insights into quantum algorithms.  For example, the maximal degree of entanglement can often be bounded in the description of tensor network state itself. In other words, the bond dimension (the dimension of the wires) in the tensor network act to bound the maximal entanglement.
Merging quantum computation with ideas from tensor networks provides new tools to quantify the entanglement that a given quantum circuit can generate \cite{Biamonte_2020, 2013JPhA...46U5301B}.

For the sake of simplicity, we work in the black-box setting and assume access to a provided unitary $Q$. The black-box setting does not consider the implementation of $Q$.  Prima facie this appears to be a limitation: in practice however, the restriction can easily be lifted.  For example, in QAOA the problem Hamiltonian can be applied for varying times, offering a natural extension of the oracle idea by giving $Q$ a simple time dependence \cite{Morales_2018}.   

In the Discussion we drop the black box access restriction and cast the steps needed to perform a meaningful near-term demonstration of our algorithm on a quantum computer, providing a low-rank approximation to eigenvectors of the quantum computers free- (or effective) Hamiltonian. The presented algorithm falls into the class of variational quantum algorithms \cite{2014NatSR...4E3589Y, 43965, 2014NatCo...5E4213P, 2017Natur.549..242K, 2019arXiv190304500B,Xia2018, 2018npjQI...4...65G,Akshay_2020}.  It returns a classical description, in the form of a tensor network, of an eigenvector of an operator found through an iterative classical-to-quantum optimization process. 

We present a general framework to determine tensor networks using quantum processors.  We focus on 1D, which enables several results related to the maximum amounts of entanglement required to demonstrate these methods.  This analysis is followed by a discussion focused on applications of these techniques, and what might be required for a meaningful near-term experimental demonstration.

\newpage

\section*{Methods}

The algorithm we propose solves the following problem: {\it given black-box access to a unitary $Q$, find any eigenvector of $Q$.}

We work in the standard mathematical setting of quantum computing.  We define $n$ qubits arranged on a line and fix the standard canonical (computational) basis. We consider the commutative Hermitian subalgrebra generated by the $n$-projectors 
\begin{equation}
    P_i = \ketbra{0}{0}_i
\end{equation}
where the subscript $i$ denotes the corresponding $i$th qubit acted on by $P_i$, with the remainder of the state-space acted on trivially. These form our observables. 

Rank is the maximum Schmidt number (the non-zero singular values) across any of the $n-1$ step-wise partitions of the qubits on a line.  Rank provides an upper-bound on the bipartite entanglement that a quantum state can support---as will be seen, a rank-$r$ state has at most $k = \log_2(r)$ ebits of entanglement. The quantum algorithm we present works by finding a sequence of maximally $k$-ebit approximations, where the $k$'th approximation can be used to seed the $(k+1)$'th approximation.  

An ebit is the amount of entanglement contained in a maximally entangled two-qubit (Bell) state. A quantum state with $q$ ebits of entanglement (quantified by any entanglement measure) has the same amount of entanglement (in that measure) as $q$ Bell states. If a task requires $l$ ebits, it can be done with $l$ or more Bell states, but not with fewer.  Maximally entangled states in 
\begin{equation}
    \mathbb{C}^d\otimes \mathbb{C}^d
\end{equation}
have $\log_2(d)$ ebits of entanglement.  The question is then to upper bound the maximum amount of entanglement a given quantum computation can generate, provided a coarse graining to classify quantum algorithms in terms of both the circuit depth, as well as the maximum ebits possible.  For low-depth circuits, these arguments are surprisingly relevant.  

We parameterize a circuit family generating matrix product states with ${\bf \theta}$\textcolor{blue}{,} a real vector with entries in $[0, 2\pi)$.  We consider action on the initial rank-1 state $\ket{\bf 0} = \ket{0}^{\otimes n}$ and define two states
\begin{equation}
    \ket{\psi(\bf \theta) } = U^\dagger ({\bf \theta})Q U ({\bf \theta})\ket{\bf 0}
\end{equation}
and 
\begin{equation}
    \ket{\tilde{\psi}(\bf \theta)} = U ({\bf \theta})\ket{\bf 0},
\end{equation}
both of yet to be specified rank. 

We will construct an objective function \eqref{eqn:log-likelihood} to minimize and hence to recover our approximate eigenvector.  The choice of this function provides a desirable degree of freedom to further tailor the algorithm to the particular quantum processor at hand. We choose  
\begin{equation}
 p_i({\bf \theta}) = \bra{\psi(\bf \theta)}P_i \ket{\psi(\bf \theta)}
\end{equation} 
and call 
\begin{equation}\label{eqn:log-likelihood}
\mathcal{L}({\bf \theta}) = \sum_{i=1}^n \ln p_i({\bf \theta}) 
\end{equation} 
the log-likelihood function of the $n$-point correlator 
\begin{equation}
    \prod_{i=1}^n p_i({\bf \theta}).
\end{equation}
The minimization of \eqref{eqn:log-likelihood} corresponds to maximizing the probability of measuring each qubit in $\ket{0}$. This minimization can be done using a variety of optimization and machine learning algorithms. The table below summarizes the steps of the algorithm.

\begin{algorithm}
\begin{algorithmic}
\STATE{Choose the maximum number of ebits $k_\text{max}$}
\STATE{Choose the maximum number of optimization iterations $n_\text{it}$}
\FOR{$k \gets 1$ to $k_\text{max}$}
    \STATE{Construct the ansatz $U_k$ corresponding to a $k$-ebit MPS}
    \STATE{Set $\boldsymbol{\theta}_k$ randomly}
    \FOR{$j \gets 1$ to $n_\text{it}$}
        \STATE{Evaluate $\boldsymbol{p}(\boldsymbol{\theta}_k)$} 
        \STATE{Evaluate $\mathcal{L}(\boldsymbol{p})$}
        \STATE{Update $\boldsymbol{\theta}_k$ using a classical optimizer}
    \ENDFOR
    \STATE{Store $\mathcal{L}_k = \mathcal{L}(\boldsymbol{p})$}
\ENDFOR
\STATE{\textbf{return} 
$\left\{\boldsymbol{\theta}_k\right\}_{k=1}^{k_\text{max}},
 \left\{\mathcal{L}_k\right\}_{k=1}^{k_\text{max}}$}.
\end{algorithmic}
\caption{Find successive tensor network approximations of an eigenvector of $Q$}
\end{algorithm}

The algorithm begins with rank-1 qubit states as  
\begin{equation}
\ket{\tilde{\psi}({\bf \theta})}=\bigotimes_{i=1}^n (\cos  \theta_1^i \ket{0} + e^{-\imath \theta_2^i}\sin \theta_1^i \ket{1}).
\end{equation}
Minimization of the objective function \eqref{eqn:log-likelihood} returns $2n$ real numbers describing a local matrix product state. Approximations of higher rank are made by utilizing the quantum circuit structure given in Figure \ref{fig:mps}. 

\begin{figure*}
 \begin{tabularx}{\textwidth}{ m{0.4\linewidth}  X  p{0.4\linewidth} }

	\centering
	\begin{circuitikz}
    	\myTriangleBlack{-0.1}{0.1}{90}{0};
  
    	\draw (1.25, 0) node[black] {$\cdots$};
  
      	\myTriangleBlack{1.9}{0.1}{90}{0};

      	\myTriangleBlack{2.9}{0.1}{90}{0};
        
        \myTriangleBlack{3.9}{0.1}{90}{0};

		\draw (0.25, -0.25) -- (0.25, -0.7);
		\draw (2.25, -0.25) -- (2.25, -0.7);
   		\draw (3.25, -0.25) -- (3.25, -2);
   		\draw (4.25, -0.25) -- (4.25, -3.5);


		\draw[fill][rounded corners=1ex] (0,-0.7) rectangle (2.5,-1.4);
		\draw (1.25, -1.05) node[white] {1};

		\draw (0.25, -1.4) -- (0.25, -5);
    
		\draw (1.4, -1.4) -- (1.4, -2);
    	\draw (1.85, -1.7) node[black] {$\cdots$};
		\draw (2.25, -1.4) -- (2.25, -2);
    	\draw (2.8, -1.7) node[black] {$\cdots$};


		\draw[fill][rounded corners=1ex] (1,-2) rectangle (3.5,-2.7);
		\draw (2.25, -2.35) node[white] {2};

		\draw (1.4, -2.7) -- (1.4, -5);
		\draw (2.25, -2.7) -- (2.25, -3.5);
    	\draw (2.8, -3.1) node[black] {$\cdots$};
		\draw (3.25, -2.7) -- (3.25, -3.5);
    	\draw (3.8, -3.1) node[black] {$\cdots$};


		\draw[fill][rounded corners=1ex] (2,-3.5) rectangle (4.5,-4.2);
		\draw (3.25, -3.85) node[white] {3};
    	\draw (5, -3.85) node[black] {$\cdots$};

		\draw (2.25, -4.2) -- (2.25, -5);
    	\draw (2.8, -4.5) node[black] {$\cdots$};
		\draw (3.25, -4.2) -- (3.25, -5);
    	\draw (3.8, -4.5) node[black] {$\cdots$};
		\draw (4.25, -4.2) -- (4.25, -5);
	\end{circuitikz}
    &
    	\centering
    	$\longrightarrow$
    &
    	\begin{tabular}{c}
        	\begin{circuitikz}
    			\draw [fill](0,0) circle [radius=0.3]  node [white] {$1$};
            	\draw (0.2, 0.2) -- (1.3, 0.2);
            	\draw [scale=0.5](1.5, 0.2) node[black] {$\vdots$};
            	\draw (0.2, -0.2) -- (1.3, -0.2);
    			\draw [fill](1.5,0) circle [radius=0.3]  node [white] {$2$};
            	\draw (1.7, 0.2) -- (2.8, 0.2);
            	\draw [scale=0.5](4.5, 0.2) node[black] {$\vdots$};
            	\draw (1.7, -0.2) -- (2.8, -0.2);
    			\draw [fill](3,0) circle [radius=0.3]  node [white] {$3$};
            	\draw (3.7, 0) node [black] {$\cdots$};
            
            	\draw(0,-0.3) -- (0, -1);
            	\draw(1.5,-0.3) -- (1.5, -1);
            	\draw(3,-0.3) -- (3, -1);
    		\end{circuitikz}
			\\
            
    		\begin{circuitikz}
            	\draw (-0.8,0) node [black] {$\downarrow$};
                \draw (0,0) node{};
            \end{circuitikz}
            \\
            
         	\begin{circuitikz}
     			\draw [fill](0,0) circle [radius=0.3]  node [white] {$1$};
             	\draw (0.3, 0) -- (1.2, 0);
     			\draw [fill](1.5,0) circle [radius=0.3]  node [white] {$2$};
             	\draw (1.8, 0) -- (2.7, 0);
     			\draw [fill](3,0) circle [radius=0.3]  node [white] {$3$};
             	\draw (3.7, 0) node [black] {$\cdots$};
            
             	\draw(0,-0.3) -- (0, -1);
             	\draw(1.5,-0.3) -- (1.5, -1);
             	\draw(3,-0.3) -- (3, -1);
     		\end{circuitikz}
       	\end{tabular}
        \\        
        &
        &         
        \\
 \end{tabularx}
 \caption{Example of a tensor network as a quantum circuit. (left) Quantum circuit realization of a matrix product state with open boundary conditions. (right) Using standard graphical rewrite rules---or by manipulating equations---one readily recovers the familiar matrix product state depiction as a `train of tensors'.}\label{fig:mps} 
\end{figure*}

\section*{Results}

The algorithm works given only oracle access to a unitary $Q$. The spectrum of $Q$ is necessarily contained on the complex unit circle and so we note immediately that 
\begin{eqnarray}
\label{eqn:max}
1 = \max_\phi |\bra{\phi}Q \ket{\phi}|^2 
&\geq& \max_{\bf \theta} |\braket{\bf 0}{\psi(\bf \theta)}|^2\textbf{}\nonumber \\
&=& \max_{\bf \theta} |\bra{\tilde{\psi}(\bf \theta)}Q\ket{\tilde{\psi}(\bf \theta)}|^2
\end{eqnarray}
with equality of the left-hand-side if and only if $\ket{\phi}$ is an eigenvector of $Q$. One advantage of the presented method is that it terminates when the measurement reaches a given value.  This implies that the system is in an eigenstate.  Such a certificate is not directly established using other variational quantum algorithms.  

Importantly, the maximization over $\bf \theta$ on the right-hand-side of \eqref{eqn:max} corresponds to the minimization of the log-likelihood \eqref{eqn:log-likelihood}.  We will then parameterize $\tilde{\psi}({\bf \theta}_k)$ where $k$ denotes a $k$-ebit matrix product state of interest.  Learning this matrix product state recovers an approximation to  an eigenvector of $Q$. With a further promise on $Q$ that all eigenvectors have a rank-$p$ matrix product state representation, then we conclude that $r<p$ implies a fundamental error in our approximation.  We consider then that the $r$'th singular value of the state takes the value $\varepsilon$. It then follows that the one-norm error scales with $O(\varepsilon)$ and the two-norm error scales only with $O(\varepsilon^2)$.
In general, we arrive at the monotonic sequence ordered by the following relation 
\begin{eqnarray}
1 &\geq& \max_{{\bf \theta}_{k+1}} |\bra{\tilde{\psi}({\bf \theta}_{k+1})}Q\ket{\tilde{\psi}({\bf \theta}_{k+1})}|^2\nonumber \\
&\geq& \max_{{\bf \theta}_{k}} |\bra{\tilde{\psi}({\bf \theta}_{k})}Q\ket{\tilde{\psi}({\bf \theta}_{k})}|^2
\end{eqnarray}
valid for $k=1$ to $\lfloor {n/2} \rfloor$ (minimum to maximum possible number of ebits).

Indeed, increasing the rank of the matrix product state approximation can improve the eigenvector approximation.  Yet it should be noted that ground state eigenvectors of physical systems are in many cases known to be well approximated with low-rank matrix product states \cite{2017arXiv170301302B, 2014AnPhy.349..117O, Vidal2010, MPSreview08, TNSreview09, 2011AnPhy.326...96S,  2014EPJB...87..280O,2013arXiv1308.3318E, 2011JSP...145..891E}.  This depends on further properties of $Q$ and is a subject of intensive study in numerical methods, further motivating the quantum algorithm we present here.  We will develop our algorithm agnostic to $Q$, leaving a more specific near-term demonstration (in which $Q$ is implemented by e.g.~a free-Hamiltonian) to the Discussion.  Generally we will express any $\ket{\tilde{\psi}({\bf \theta})}$ as a matrix product state as
\begin{equation}\label{eqn:mps}
\ket{\tilde{\psi}({\bf \theta})} = \sum_{q, s,\dots,n} A_q^{[{\bf \theta}_q]} 
A_s^{[{\bf \theta}_s]} \cdots A_n^{[{\bf \theta}_n]} \ket{q, s, \dots, n}.
\end{equation}
In \eqref{eqn:mps} the rank $r$ of the representation is embedded into the realization of the $A$'s.  Quantum mechanics allows the deterministic generation of a class of isometries, where an isometry $U$ that is also an endomorphism on a particular space is called unitary. 

Matrix product states \eqref{eqn:mps} are not isometries---though correlation functions are readily calculated from them. 
Furthermore, matrix product states can be deterministically generated by the uniform quantum circuit given in Figure~\ref{fig:mps}.
Other isometric structures of interest include trees and so-called, Multiscale Entanglement Renormalization Ansatz (MERA) networks~\cite{Vidal2010, 2007PhRvL..99v0405V, 2008PhRvL.101r0503G, 2008PhRvL.101k0501V}.

Consider then a rank-$r$ approximation to an eigenvector of $Q$.  The blocks in Figure \ref{fig:mps} represent unitary maps.  These circuits act on at most $\lceil \log_2(r)\rceil$ qubits.  Hence, each of these blocks has at most $r^2$ real degrees of freedom in $[0, 2\pi)$.  The general realization of these blocks using the typical basis of CNOT gates and arbitrary local unitaries can be done by a range of methods, see i.e.~\cite{2004PhRvL..93m0502M}.  A commonly used theoretical lower bound requires 
\begin{equation}
\frac{1}{4}(r^2 - 3\log_2 r - 1)
\end{equation}
CNOT gates, where the method in \cite{2004PhRvL..93m0502M} requires $r^2$ local qubit gates and did not reach this theoretical lower bound of CNOT gates.  The total number of single qubit and CNOT gates nevertheless scales as $O(r^2)$ for each block, where the number of blocks is bounded by $n$. Hence the implementation complexity scales as $O(l\cdot n\cdot r^2)$, where the optimization routine terminates after $l$ steps (perhaps in a local minimum).  

Instead of preparing $\ket{\tilde{\psi}({\boldsymbol \theta})}$ by  a quantum circuit with ${\boldsymbol \theta} \in (0, 2\pi]^{\times l}$ tunable parameters as 
\begin{equation}
\ket{\tilde{\psi}({\boldsymbol \theta})} = \prod _lU_l \ket{0}^{\otimes n}
\end{equation}
where $U_l$ is adjusted by $\theta_l$ one might adopt an alternative (heuristic) circuit realization preformed by adjusting controllable Hamiltonian parameters realizing each block, subject again to minimization of \eqref{eqn:log-likelihood}.  With such an approach, one will prepare $\ket{\tilde{\psi}({\boldsymbol \theta})}$ by tuning accessible time-dependent parameters ($\theta_k(t)$ corresponding to Hermitian $A^k$) as 
\begin{equation}
\ket{\tilde{\psi}} =\mathcal{T}\{ e^{-\imath \sum \theta_k(t) A^k}\}\ket{0}^{\otimes n}
\end{equation} 
where $\mathcal{T}$ time orders the sequence and superscript $k$ indexes the $k$th operator $A^k$. Provided these sequences are localized appropriately, the matrix product structure still remains.

We then consider vertical partitions of a quantum circuit with the $n$ qubits positioned horizontally on a line.  For an $m$-depth quantum circuit (where $m$ is presumably bounded above by a low-order polynomial in $n$), the maximum number of two-qubit gates crossed in a vertical partition is never more than $m$.  
The maximum number of ebits generated by a fully entangling two-qubit CNOT gate is never more than a single ebit.
We then consider the $(n-1)$ partitions of the qubits, the maximum partition with represent to ebits is into two (ideally) equal halves, which is never more than $\lceil n/2 \rceil$. 
We then arrive at the general result that an $m$-depth quantum circuit on $n$ qubits never applies more than
\begin{equation}
    \min\{\lceil n/2 \rceil, m\}
\end{equation}
ebits of entanglement.  This immediately puts a lower-bound of $\sim n/2$ on the two-qubit gate-depth for $Q$ to potentially drive a system into a state supporting the maximum possible ebits of entanglement.

\begin{figure}
 \centering
 \includegraphics[width=.45\textwidth]{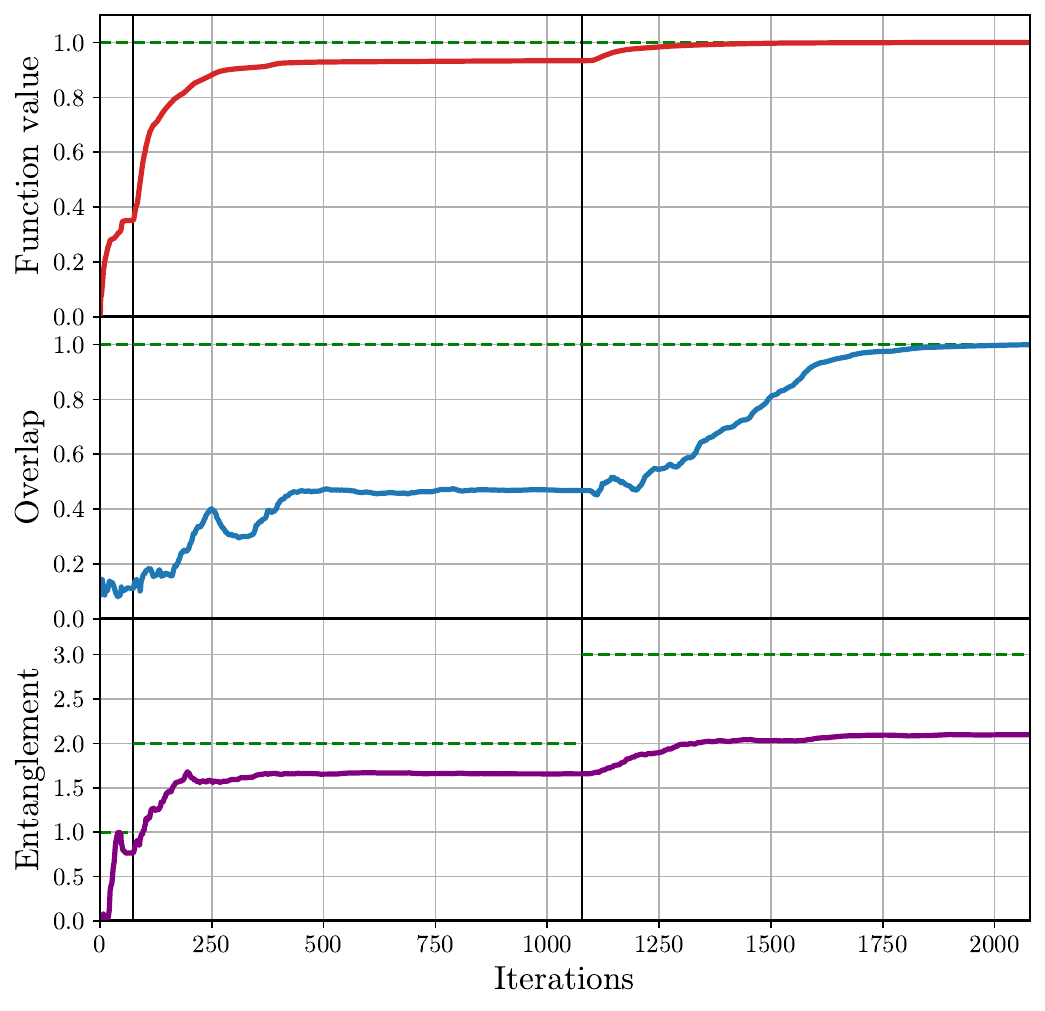}
 \caption{Algorithm demonstration on a randomly generated 6-qubit unitary $Q$: the value of \eqref{eqn:max} (upper), overlap between the variational state $\ket{\tilde{\psi}({\bf \theta})}$ and the closest eigenstate of $Q$ (middle) and the von Neumann entropy of the subsystem of the first three qubits (lower). The vertical black solid lines correspond to $k$: the points between $1$ and $75$ iterations are obtained with $1$-ebit MPS, $76$-$1075$ iterations with $2$-ebit MPS, and $1076$-$2075$ iterations with $3$-ebit MPS.}
 \label{fig:plots} 
\end{figure}

In Figure \ref{fig:plots} we demonstrate our algorithm on finding an eigenstate of a randomly generated 6-qubit unitary matrix.
For minimizing the function \eqref{eqn:log-likelihood}, we used the Broyden-Fletcher-Goldfarb-Shanno (BFGS) minimization method \cite{NoceWrig06}.
For each $k$-ebit MPS, we place the $k$-layered hardware-efficient ansatz as the operators in blocks \cite{2017Natur.549..242K}.
The number of optimization iterations $n_\text{it}$ is set to $1000$.

\section*{Discussion}

We now turn to the realization of $Q$ and sketch a possible demonstration for a near-term device.  Polynomial time simulation of Hamiltonian evolution is well known to be {\bf BQP}-hard.  This provides an avenue for $Q$ to represent a problem of significant computational interest, as simulating quantum evolution and quantum factoring are in {\bf BQP}.  We aim to bootstrap properties of the quantum processor as much as possible to reduce resources for a realization---see for example \cite{2017Natur.549..242K}.  

Let $Q(t)$ be the one-parameter unitary group generated by $\mathcal{H}$, where $\mathcal{H}$ represent\textcolor{blue}{s} a 3-SAT instances.  Given access to an oracle computing  
\begin{equation}
    \bra{\tilde{\psi}({\bf \theta}_{1})}\mathcal{H}\ket{\tilde{\psi}({\bf \theta}_{1})},
\end{equation}
we can minimize over all eigenvectors, which is {\bf NP}-hard.  Hence, finding even rank-$1$ states can be {\bf NP}-hard.  This provides a connection between our method and QAOA \cite{Fahri-qaoa1}. Similarly, we can also use this external minimization to connect our method to VQE \cite{2014NatCo...5E4213P}.  However, our method provides a certificate that on proper termination, the system is indeed in such a desired eigenstate.  

When $\mathcal{H}$ is a general quantum Hamiltonian, minimization of
\begin{equation}
    \bra{\tilde{\psi}({\bf \theta}_{k})}\mathcal{H}\ket{\tilde{\psi}({\bf \theta}_{k})}
\end{equation}
is in turn, {\bf QMA}-hard.  For example, pairing our procedure with an additional procedure (quantum phase estimation) to minimize $Q$ over all eigenvectors would hence provide rank-$k$ variational states and hence our methods provide a research direction which incorporates tensor network methods in works such as  e.g.~\cite{43965, 2014NatCo...5E4213P, 2017Natur.549..242K}.  It should however be noted that phase estimation adds significant experimental difficultly compared with the algorithm presented here and the algorithm is closer to VQE (with evident differences as listed above and in the main text).  

For a near-term demonstration, we envision $Q$ to be realized by bootstrapping the underlying physics of the system realizing $Q$, e.g.~using the hardware efficient ansatz \cite{2017Natur.549..242K}. For instance, one can realize $Q$ as a modification of the systems free-Hamiltonian using effective Hamiltonian methods (modulating local gates).  This greatly reduces practical requirements on $Q$. 

The interaction graph of the Hamiltonian generating $Q$ can be used to define a PEPS tensor network  (as it will have the same structure as the layout of the chip, it will no longer have the contractable properties of matrix product states yet is still of interest) \cite{MPSreview08}.
The algorithm works otherwise unchanged, but the circuit acts on this interaction graph (instead of a line) to create a corresponding tensor network state (a quantum circuit in the form of e.g.~the variational ansatz).  Tailored free-evolution of the system Hamiltonian generates $Q$.  Our algorithm returns a tensor network approximation of an eigenstate of $Q$.

The first interesting demonstrations of the quantum algorithm we have presented should realize rank-$2$ tensor networks (matrix product state), and the corresponding tensor network can be realized with a few hundred gates for a system with a few hundred qubits.  

\begin{acknowledgments}
A.K.~and J.B.~acknowledge support from agreement No.~014/20, {\it Leading Research Center on Quantum Computing}.
A.U.~acknowledges RFBR Project No.~19-31-90159 for financial support. This manuscript has been released as a preprint as \url{arXiv:1804.02398} \cite{biamonte2018quantum}.
\end{acknowledgments}

\bibliographystyle{unsrt}
\bibliography{bibliography}

\end{document}